\documentclass[aps,superscriptaddress,twocolumn,prr]{revtex4-1}
\usepackage{amssymb,amsmath,mathptmx}
\usepackage{graphicx}
\usepackage{dcolumn}
\usepackage{bm}
\usepackage{hyperref}
\usepackage{color}
\usepackage[utf8x]{inputenc}
\usepackage{array}
\usepackage{ulem}
\newcolumntype{C}{>{\centering\arraybackslash}p{1em}}

\newcommand{\be}{\begin{equation}}
\newcommand{\ee}{\end{equation}}
\newcommand{\bea}{\begin{eqnarray}}
\newcommand{\eea}{\end{eqnarray}}
\begin{document}

\title{Proposal for detecting the $\pi-$shifted Cooper quartet supercurrent}

\author{R\'egis M\'elin}
\email{regis.melin@neel.cnrs.fr}

\affiliation{Univ. Grenoble-Alpes, CNRS, Grenoble INP\thanks{Institute
    of Engineering Univ. Grenoble Alpes}, Institut NEEL, 38000
  Grenoble, France}

\author{Romain Danneau}

\affiliation{Institute for Quantum Materials and Technologies, Karlsruhe Institute of Technology,
Karlsruhe D-76021, Germany}

\author{Clemens B. Winkelmann}

\affiliation{Univ. Grenoble-Alpes, CNRS, Grenoble INP\thanks{Institute
    of Engineering Univ. Grenoble Alpes}, Institut NEEL, 38000
  Grenoble, France}

\begin{abstract}
The multiterminal Josephson effect aroused considerable interest
recently, in connection with theoretical and experimental evidence for
correlations among Cooper pairs, that is, the so-called Cooper
quartets.  It was further predicted that the spectrum of Andreev bound
states in such devices could host Weyl-point singularities. However,
the relative phase between the Cooper pair and quartet supercurrents
has not yet been addressed experimentally. Here, we propose an
experiment involving four-terminal Josephson junctions with two
independent orthogonal supercurrents, and calculate the critical
current contours (CCCs) from a multiterminal Josephson junction
circuit theory. We predict a generically $\pi$-shifted contribution of both the 
local or nonlocal second-order
  Josephson harmonics. Furthermore, we show that these lead to marked
   nonconvex shapes for the CCCs in zero
magnetic field, where the dissipative state reenters into the
superconducting one. Eventually, we discuss distinctive features of the non-local Josephson processes in the CCC's. The experimental observation of the latter could
allow providing firm evidence of the $\pi$-shifted Cooper
quartet current-phase relation.
\end{abstract}

\maketitle
\section{Introduction}

Entanglement in electronic superconducting circuits is central to
quantum engineering, and prototypes of quantum processors were
recently realized, unveiling a variety of physical
phenomena~\cite{Google1}.  Entanglement engines were proposed in the
early 2000s, with normal metal-superconductor-normal metal
($N$-$S$-$N$) hybrids as sources of entangled Einstein-Podolsky-Rosen
pairs of electrons
\cite{theory-noise2,theory-CPBS1,theory-CPBS4,theory-CPBS6,theory-CPBS7,theory-CPBS8,theory-CPBS11,Yeyati2007,theory-CPBS7,theory-noise8,theory-noise9,theory-noise11}.
A series of experiments addressed nonlocality in the DC current
response
\cite{exp-CPBS1,exp-CPBS2,exp-CPBS3,exp-CPBS6,exp-CPBS5,Tan2015,Borzenets2016a,Danneau}
and quantum noise \cite{exp-CPBS8} as evidence for entangled split
Cooper pairs
\cite{theory-noise2,theory-CPBS1,theory-CPBS4,theory-CPBS6,theory-CPBS7,theory-CPBS8,theory-CPBS11,Yeyati2007,theory-CPBS7,theory-noise8,theory-noise9,theory-noise11}.
On the other hand, the emerging field of all-superconducting
multiterminal Josephson junctions
\cite{1979,Omelyanchouk1,Omelyanchouk2,Omelyanchouk3,Omelyanchouk4}
offers new perspectives such as exotic transient quantum correlations
among Cooper pairs, known as Cooper quartets
\cite{Cuevas-Pothier,Freyn,Melin1,Jonckheere,Akh2,MF,split-quartets}. While
a series of experiments reported clear signatures of Cooper
quartets~\cite{Lefloch,Heiblum,HGE,Pribiag}, these features were not
observed by
others~\cite{exp-multiter-supp,multiterminal-exp2,multiterminal-exp3,multiterminal-exp4,multiterminal-exp5,multiterminal-exp6,multiterminal-exp9,multiterminal-exp10},
possibly due to delicate material and device fabrication issues. In
parallel, multiterminal Josephson junctions also focused strong
interest recently as a testbed of Floquet theory
\cite{FWS,engineering,Berry,paperII,paperIII,ultralong,Keleri,Park},
as well as a platform for the emergence of energy level repulsion in
Andreev molecules
\cite{Pillet,Pillet2,Nazarov-PRR,multiterminal-exp7,IBM,Tarucha,Tarucha2},
the production of Weyl-point singularities in the Andreev spectrum
\cite{vanHeck,Padurariu,multiterminal-exp1,Nazarov1,Nazarov2,topo0,topo2,topo3,topo1-plus-Floquet,topo1,topo4,topo4-bis,topo5,Berry,Feinberg1,Feinberg2,Levchenko1,Levchenko2,Akh2,Gavensky},
and the multiterminal superconducting diode effect~\cite{SDE1,SDE2}.

In spite of intense experimental efforts for observing signatures of
the quartet state and its new physics beyond the standard Resistively
Shunted Josephson Junction model \cite{Lefloch,HGE,Heiblum}, novel
schemes are necessary for ascertaining the Cooper quartets.  When
driving current between pairs of contacts in a multiterminal Josephson
junction with an even number $2n$ of superconducting leads, $n$
equations of current conservation are imposed by the external
circuit. Those $n$ constraints (for a total of $2n$ phase variables)
allow for supercurrent inside a region in phase space parameterized by
$2n-n\equiv n$ independent variables. With four terminals, a DC
supercurrent is thus established within a two-dimensional region in
the plane of the bias currents, separated from the resistive state by
a one-dimensional critical current contour (CCC). In a recent work,
Pankratova {\it et al.}  \cite{multiterminal-exp3} reported nonconvex
shapes in the CCCs of four-terminal semiconductor-superconductor
Josephson junctions. However, these nontrivial features appeared only
at rather high magnetic fields, corresponding to about half a flux
quantum threading the central part of the device. The observation of
nonconvex CCCs was interpreted using Random Matrix Theory, assuming
time-reversal symmetry breaking, either due to an applied magnetic
field or preexisting in the normal state \cite{multiterminal-exp3}.
  
\begin{figure}[t]
  \centerline{\includegraphics[width=\columnwidth]{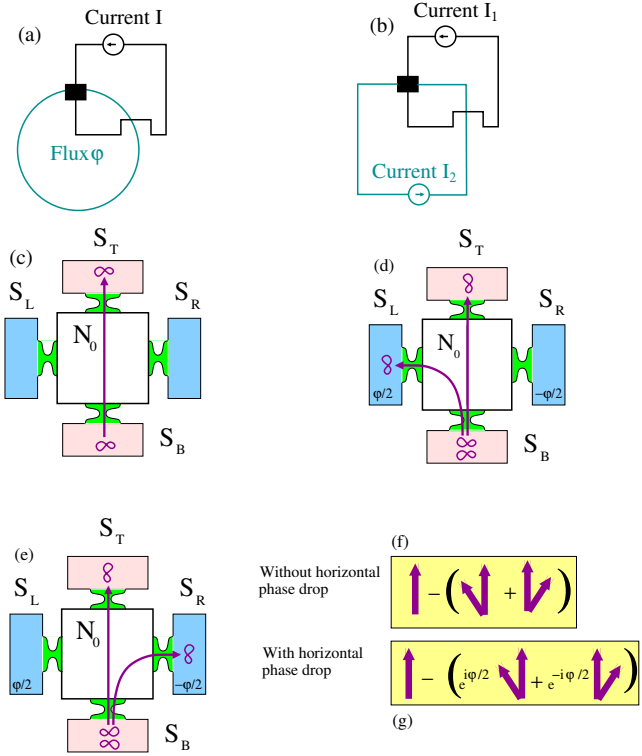}}
  \caption{Sketch of the superconducting four-terminal device with
    either one current and one phase bias (a), or two orthogonal
    current biases (b). The superconductors $S_L$, $S_R$, $S_T$ and
    $S_B$ are connected to the normal metallic region $N_0$. The four
    $N_0$-$S_{i}$ junctions consist of tunable quantum point contacts,
    where the transmission of the $N_0$-$S_T$ interface is reduced by
    a scaling factor $\tau_T$. Panels c-e represent the lowest-order
    Josephson processes occurring in a simplified toy-model. Panel c
    shows the two-terminal DC-Josephson effect from $S_B$ to $S_T$,
    which is insensitive to the horizontal contacts. Panels d and e
    show the Cooper quartet processes, which take two Cooper pairs
    from $S_B$, exchange partners and transmit the outgoing pairs into
    $(S_T,\,S_L)$ and $(S_T,\,S_R)$, respectively. In presence of a
    horizontal phase drop, these two processes pick up opposite
    phases, as shown in panels f and g. This leads to interfering
    quartet supercurrent components within this simplified model,
    without and with horizontal phase drop, respectively. Due to the
    $\pi$-shift, the critical current along the vertical direction in
    panel f is reduced by the two quartet processes. On panel g, a
    phase drop along the horizontal direction dephases the negative
    contribution of both processes, resulting in an increased critical
    current and thus a nonconvex CCC.
    \label{fig:thedevice}
  }
\end{figure}

Here, we demonstrate that in the presence of at least one contact with
an intermediate transmission, another mechanism for the emergence of
nonconvex CCCs is possible, which does not require a magnetic
field. Namely, we find correspondence between the {\it quartet
  physics} and the emergence of nonconvex sharp-angled points
  in the CCCs at zero magnetic field. This distinctive signature stems from the interference between
symmetric quartet channels, which are dephased by a transverse
supercurrent (see Fig.~\ref{fig:thedevice}). In
other words, we demonstrate that {\it macroscopic} critical current
measurements can probe the {\it microscopic} internal structure of
entangled split Cooper
pairs~\cite{theory-noise2,theory-CPBS1,theory-CPBS4,theory-CPBS6,theory-CPBS7,theory-CPBS8,theory-CPBS11,Yeyati2007}.

The article is organized as follows. The $\pi$-shifted quartets are
introduced in Sec.~\ref{sec:pi-shift}. The device and the model are
presented in Sec.~\ref{sec:thedevice}. The numerical results,
analytical and numerical, are presented and discussed in
Sec.~\ref{sec:theresults}. Concluding remarks are provided in
Sec.~\ref{sec:conclusions}.

\section{$\pi$-shifted Cooper quartets}
\label{sec:pi-shift}

In this section, we provide physical arguments supporting the
$\pi$-shifted Cooper quartet current-phase relation. The key
underlying concept can readily be understood starting from a
three-terminal configuration of Josephson junctions in the DC
superconducting state, connecting the leads $S$ with respective
indices $i,j,k$ \cite{Freyn,Pillet}, and biased with respective phases
$\varphi$. The corresponding spin-singlet wave-function of a split
Cooper pair for instance between $S_i$ and $S_j$ takes the form
\begin{equation}
    \label{eq:single-split}
    \psi=
    \frac{1}{\sqrt{2}}\left(c_{i,\uparrow}^+ c_{j,\downarrow}^+ -
c_{i,\downarrow}^+ c_{j,\uparrow}^+\right)
,
\end{equation}
where $c^+_{i,\sigma}$ creates a spin-$\sigma$ fermion in $S_i$.  The
splitting event in Eq.~(\ref{eq:single-split}) can come along with a
second one. The resulting composite four-fermion transient state,
i.e. a Cooper quartet \cite{Freyn,Lefloch,Heiblum,HGE,Pribiag}, ends
up as two Cooper pairs transmitted into $S_i$ and $S_j$, respectively,
and described by
\begin{equation}
  \label{eq:double-split}
\langle \psi^2 \rangle=-\langle c_{i,\uparrow}^+
c_{i,\downarrow}^+ \rangle \, \langle c_{j,\uparrow}^+ c_{j,\downarrow}^+
\rangle
,
\end{equation}
where $\langle\,...\,\rangle$ is a quantum mechanical expectation
value (details can be found in Appendix~\ref{sec:I}).  By {\it probing
  the internal structure of (double) split Cooper pairs}, we mean
providing experimental evidence for the negative sign in
Eq.~(\ref{eq:double-split}), which is a direct consequence of both
quantum mechanical exchange and the split Cooper pair structure of
Eq.~(\ref{eq:single-split}). Consequently, the relation between the
quartet supercurrent $I_q$ and the quartet phase $\varphi_q$ is
inverted:
\begin{eqnarray}
\label{eq:Iq-1}
  I_q(\varphi_q)&=&-|I^{c,q}| \sin
  \varphi_q\\
  \label{eq:Iq-1-bis}
  \varphi_q&=&\varphi_a+\varphi_b-2\varphi_c
  ,
\end{eqnarray}
where $\varphi_a$, $\varphi_b$ and $\varphi_c$ are the superconducting
phase variables of the leads $S_a$, $S_b$ and $S_c$ respectively.
Eq.~(\ref{eq:Iq-1}) can be rewritten as $I_q(\varphi_q)=|I^{c,q}| \sin
(\varphi_q+\pi)$ and this $\pi$-shift is a macroscopic signature for
the specific internal structure of single split Cooper pairs, see
Eq.~(\ref{eq:single-split}).

Another simple perspective on the $\pi$-shift of the quartets readily
follows from considering a single two-terminal superconducting weak
link with normal-state transmission $\alpha$. Here, the energy-phase
relation can be Fourier-expanded as $E^J(\varphi)=E^J_0+E^J_{2e}
\cos\varphi +E^J_{4e} \cos2\varphi+...$. The $\cos \varphi$ term
represents the Josephson Cooper-pair energy, and is dominant in the
limit of small transparency, while the $\cos 2\varphi$ one describes
correlated tunneling of two Cooper pairs. We find $E^J_{4e}/E^J_{2e}
\approx - \alpha / 16$ in the small-$\alpha$ limit and more generally
$E^J_{4e}/E^J_{2e} <0$ for all $\alpha<1$, see
Appendix~\ref{sec:II}. This negative sign echoes the above
current-phase relation of the quartets. More generally, our work
proposes a method to directly reveal these $\pi$-shifted second-order
Josephson harmonics, using a multiterminal configuration.

\section{The device and multiterminal Josephson circuit theory}
\label{sec:thedevice}
In this section, we present the two types of devices and the
approximations sustaining multiterminal Josephson circuit theory. The
proposed device consists of four BCS superconducting leads
$S_L,\,S_R,\,S_B$ and $S_T$, with the respective superconducting phase
variables $\varphi_L,\,\varphi_R,\,\varphi_B$ and $\varphi_T$, and
connected via a square-shaped normal conductor $N_0$ as shown in
Fig.~\ref{fig:thedevice}. The external circuit imposes current in
orthogonal directions, that is, a vertical current $I_{v}\equiv
I_T=-I_B$ and a horizontal one $I_{h}\equiv I_R=-I_L$. The absence of
coupling between $I_v$ and $I_h$ produces a square or rectangular CCC,
while rounded CCCs are indicative of coupling.

Our main result is that assuming a single or two contacts with
transparency smaller than the others, nonconvex CCCs emerge in the
$(I_{v},\,I_{h})$ plane already under zero applied magnetic field. We
thus find reentrance of the dissipative state into the superconducting
region as a distinctive signature of the $\pi$-shifted contribution of second-order Josephson harmonics.
Furthermore, we show that the $\pi$-shifted Cooper
 quartet supercurrent produces distinctive sharp reentrant sharp-angled points in the CCCs.

The four-terminal geometry is found by a straightforward
generalization of Josephson circuits, where now the $I_{v}$ and
$I_{h}$ supercurrents result from an interference between multipair
processes involving the phases of more than two terminals
\cite{Freyn,split-quartets}. For instance, in a two-terminal Josephson
junction, the terms corresponding to Cooper pairs transmitted from
$S_{i}$ to $S_{j}$ couple to the difference $\delta_{i,j}=
\varphi_{i}-\varphi_{j}$. Similarly, with four terminals, the relevant
phase variables are then given by gauge-invariant combinations such as
$\delta_{i,j}+\delta_{k,l}$ \cite{split-quartets}, which reduces to
Eq. (\ref{eq:Iq-1-bis}) for three terminals \cite{Freyn}.

In our multiterminal Josephson circuit model, we assume tunable
contacts with a few transmission modes connecting the four
superconductors to a central normal metal island
(see Fig.~\ref{fig:thedevice}), as was recently
demonstrated in bilayer graphene-based two-terminal Josephson
devices~\cite{Danneau2} and in multiterminal
semiconducting-superconducting quantum point contacts
\cite{Pribiag}. Considering intermediate contact transparencies,
although the DC-Josephson effect is dominant, the next-order Cooper
quartets still yield a sizable contribution, while the even
higher-order terms are smaller. This {\it hierarchy} justifies the
approach of the Letter, considering within a single four-terminal
device all the Josephson processes involving two, three and then four
terminals.  The calculation involves two steps: our starting points
are the approximate analytical expressions of the current-phase
relations discussed above, with sign and amplitude as free parameters.
This allows comparing the CCCs with respectively positive or negative
Cooper quartet contributions.  From this we will arrive to the
conclusion that nonconvex CCCs in zero field carry the unique
signature of the microscopic $\pi$-shifted Cooper quartet
current-phase relation, and would be absent with a $0$-shift.

We consider intermediate transparency interfaces, with hopping amplitudes
$J_L,\,J_R,\,J_B$ and $J_T$ connecting respectively the four superconducting leads
$S_L$, $S_R$, $S_B$ and $S_T$ to a normal tight-binding lattice
$N_0$. The DC-Josephson supercurrent of Cooper pairs from lead $S_i$
to lead $S_j$ is written as $I_{P} = I^{c,P}_{i,j}
\sin\delta_{i,j}$. The {\it nonlocal} DC-Josephson supercurrent
of the Cooper quartets  
involves, at the lowest order in tunneling, the following three terms:
\begin{eqnarray}
&&I_{q}=
I^{c,q}_{i,j,(k)} \sin(\delta_{i,k}+\delta_{j,k})\\& +&
\nonumber
I^{c,q}_{i,(j),k} \sin(\delta_{i,j}+\delta_{k,j}) +
I^{c,q}_{(i),j,k} \sin(\delta_{j,i}+\delta_{k,i})
\nonumber
.
\end{eqnarray}
Here, $I^{c,q}_{i,j,(k)}$ for instance represents the critical quartet
current of two pairs emitted by $S_k$ and recombining into $S_i$ and
$S_j$.  We introduce the individual channel transmissions $\tau_i$
such that all $J_i=\sqrt{\tau_i} J^{(0)}$, with $J^{(0)}$ a constant
smaller than the band-width $W$.  The critical currents scale as
follows: $I^{c,P}_{i,j}=\tau_i\tau_j I^{c(0)}_{i,j}$ for the Cooper
pairs, and $I^{c,q}_{i,j,(k)}=\tau_i\tau_j\tau_k^2
I^{c(0)}_{i,j,(k)}$, $I^{c,q}_{i,(j),k} = \tau_i\tau_j^2\tau_k
I^{c(0)}_{i,(j),k}$ and $I^{c,q}_{(i),j,k}=\tau_i^2\tau_j\tau_k
I^{c(0)}_{(i),j,k}$ for the Cooper quartets, where the $I^{c(0)}$s do
not scale with the transmissions.

\section{Results}
\label{sec:theresults}

\subsection{Polarization with one current and one phase bias}
\label{sec:mixt-pol}

In this subsection we present analytical results for the device
polarized with one current and one phase bias, see
Fig.~\ref{fig:thedevice}a. An external source drives a supercurrent
from $S_B$ to $S_T$ and an external loop fixes the phase difference
between $S_L$ and $S_R$. We additionally assume that the $N_0$-$S_T$
link has a tunneling amplitude $J_T$ small compared to
$J_L=J_R=J_B\equiv J^{(0)}$, i.e. $\tau_T\alt
\tau_L,\,\tau_R,\,\tau_B\alt 1$. Then, we make a perturbation
expansion in tunneling of the Josephson circuit to the dominant order
$\tau_T$, neglecting the processes of order $\tau_T^2$ (see
Appendix~\ref{sec:III}). In absence of the quartets, we find two types
of processes: (i) The direct two-terminal DC-Josephson effect of the
Cooper pairs from $S_B$ to $S_T$ (see Fig.~\ref{fig:thedevice}e), and
(ii) The two-terminal DC-Josephson processes of the Cooper pairs
involving the lateral superconductors $S_L$ and $S_R$. Adding now the
quartets, we include all possible processes appearing at the orders
$\tau_T^0$ and $\tau_T$.

The cartoon shown in Fig.~\ref{fig:thedevice}
illustrates the case where, at the order $\tau_T$, the critical
current $I^c_{v}$ from $S_B$ to $S_T$ results from an interference
between the amplitudes of the two-terminal DC Josephson effect and
both Cooper quartets (see Figs.~\ref{fig:thedevice}f,g). Taking an
opposite relative sign of the two- and three-terminal contributions,
respectively, leads to a reduction of $I^c_{v}$ upon including the
Cooper quartets. Notably, because each quartet process picks up an
opposite phase $\varphi_L= -\varphi_R\equiv -\varphi/2$, their
respective contributions are dephased and the value of $I^c_{v}$ is
restored upon applying a supercurrent $I_{h}$ (or a phase gradient) in
the transverse direction, as shown in Figs.~\ref{fig:thedevice}f
and~g.

\begin{figure}[htb]
  \includegraphics[width=\columnwidth]{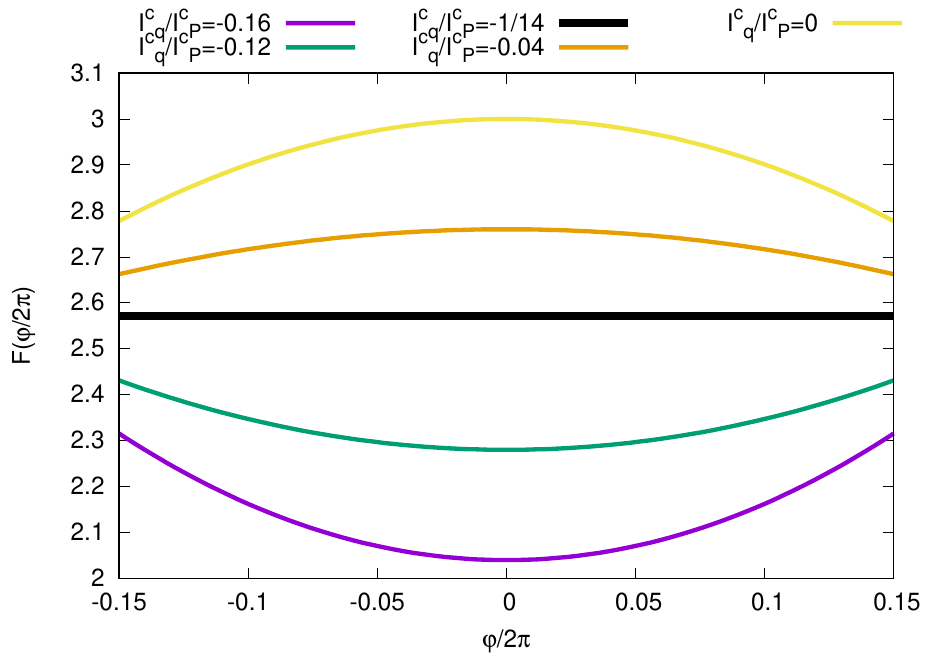}
  \caption{The figure shows the shape of the CCC,
    i.e. $F(\varphi/2\pi)$ as a function of $\varphi/2\pi$, where
    $F(\varphi/2\pi)$ is proportional to the critical current, see
    Eq.~(\ref{eq:F}). Polarization is with one current and one phase
    bias, see Fig.~\ref{fig:thedevice}a. The
    notation $\varphi$ stands for the phase difference between the
    $N_0-S_R$ and $N_0-S_L$ contacts. Two regimes are obtained if
    $I_q^c<-I_P^c/14$ or $I_q^c>-I_P^c/14$, corresponding to nonconvex
    or convex CCCs respectively.
    \label{fig:F}
  }
\end{figure}

Now, we evaluate the full set of microscopic two- and three-terminal
processes at the relevant orders (details in
Appendix~\ref{sec:III}). Using the notations $\varphi_L= -\varphi_R=
-\varphi/2$, we demonstrate in Appendix~\ref{sec:III} that, at small
$\tau_T$ and quartet Josephson energy $E_q=(\hbar/2e)I^c_q$, the
critical current $I_{v}^c$ from $S_B$ to $S_T$ can be approximated as
  \begin{equation}
    \label{eq:result}
    I_{v}^c\simeq \tau_T I_P^c
    \left\{3+6\frac{I_q^c}{I_P^c} -
\frac{\varphi^2}{4} \left[1+14\frac{I_q^c}{I_P^c}\right]\right\} ,
\end{equation}
where $I_P^c$ and $I_q^c$ are proportional to the critical currents of the two-
and three-terminal Cooper pair and Cooper quartet processes
respectively. Eventually, Eq.~(\ref{eq:result}) predicts nonconvex
CCCs if the condition
\begin{equation}
  \label{eq:<}
  I_q^c<-\frac{I_P^c}{14}
\end{equation}
is fulfilled. In this case, the dissipative state reenters into the
superconducting one, as a result of the $\pi$-shifted Cooper quartet
current-phase relation coming from the spin-singlet minus signs in
Eqs.~(\ref{eq:single-split}) and~(\ref{eq:double-split}).

We rewrite Eq.~(\ref{eq:result}) as $I_{v}^c(\varphi)= \tau_T I_P^c
F(\varphi/2\pi)$, with
\begin{equation}
  \label{eq:F}
  F\left(\frac{\varphi}{2\pi}\right)=3+6\frac{I_q^c}{I_P^c} - \pi^2
    \left(\frac{\varphi}{2\pi}\right)^2 \left[
      1+14\frac{I_q^c}{I_P^c}\right].
\end{equation}
The variations of $F(\varphi/2\pi)$ are shown in
Fig.~\ref{fig:F}, confirming emergence of nonconvex
or convex CCC if $I_q^c<-I_P^c/14$ or $I_q^c>-I_P^c/14$ respectively.

\begin{figure}[htb]
  \includegraphics[width=\columnwidth]{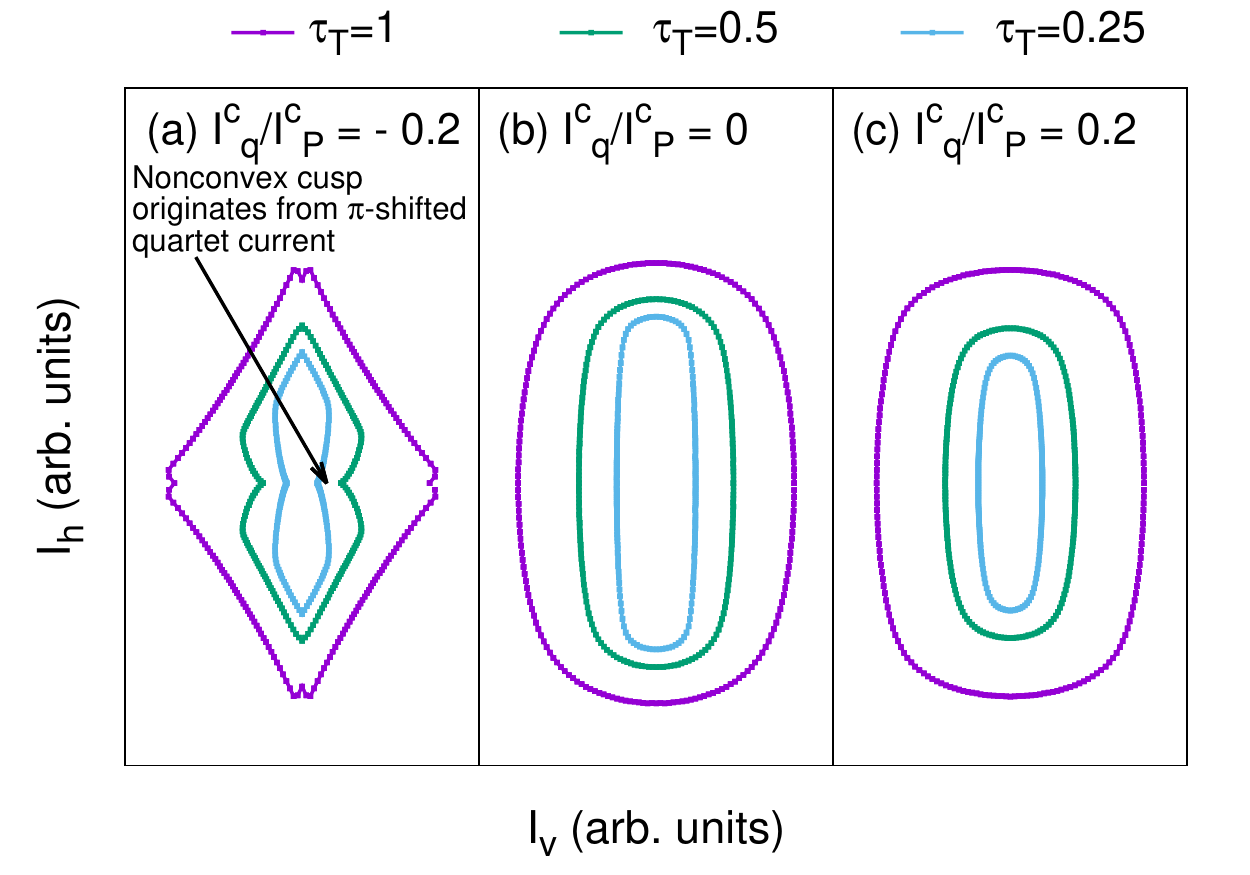}
  \caption{Critical current contours in the $(I_v,I_h)$ plane, with
    $I_q^c/I_P^c=-0.2,\,0,\,0.2$ (on panels a, b, c respectively), and
    with a single weak link. The contact transmission coefficients are
    such that $\tau_B=\tau_L=\tau_R=1$ and $\tau_T=1$ (magenta),
    $\tau_T=0.5$ (green) and $\tau_T=0.25$ (blue). Each panel is
    rescaled to full size on the $I_{v}$ and
    $I_{h}$-axis. Polarization is with two orthogonal current biases,
    see Fig.~\ref{fig:thedevice}b. Temperature is set to zero.
  \label{fig:CCC1}
 }
\end{figure}

\begin{figure*}[htb]
  \centerline{\includegraphics[width=.5\textwidth]{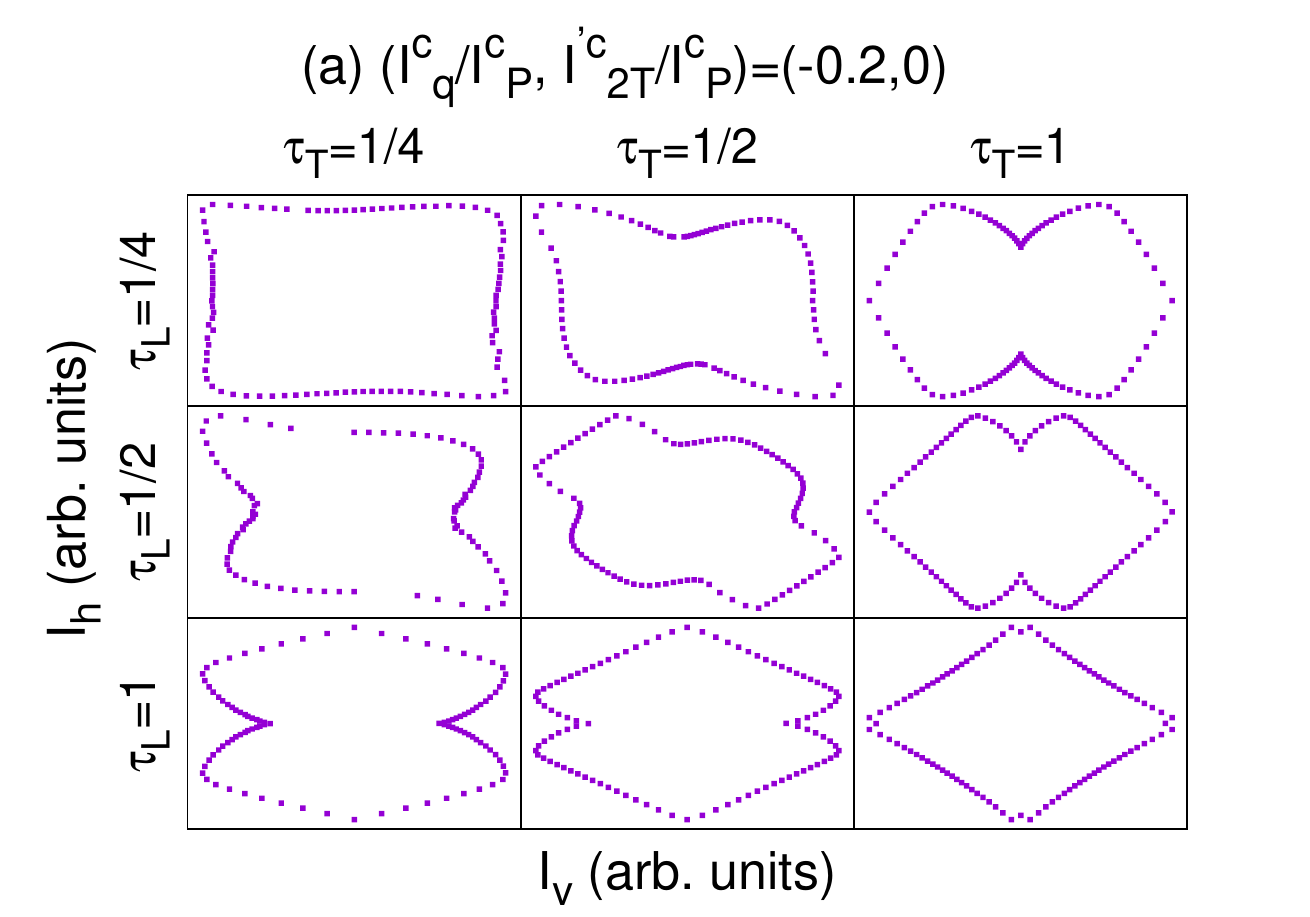}\includegraphics[width=.5\textwidth]{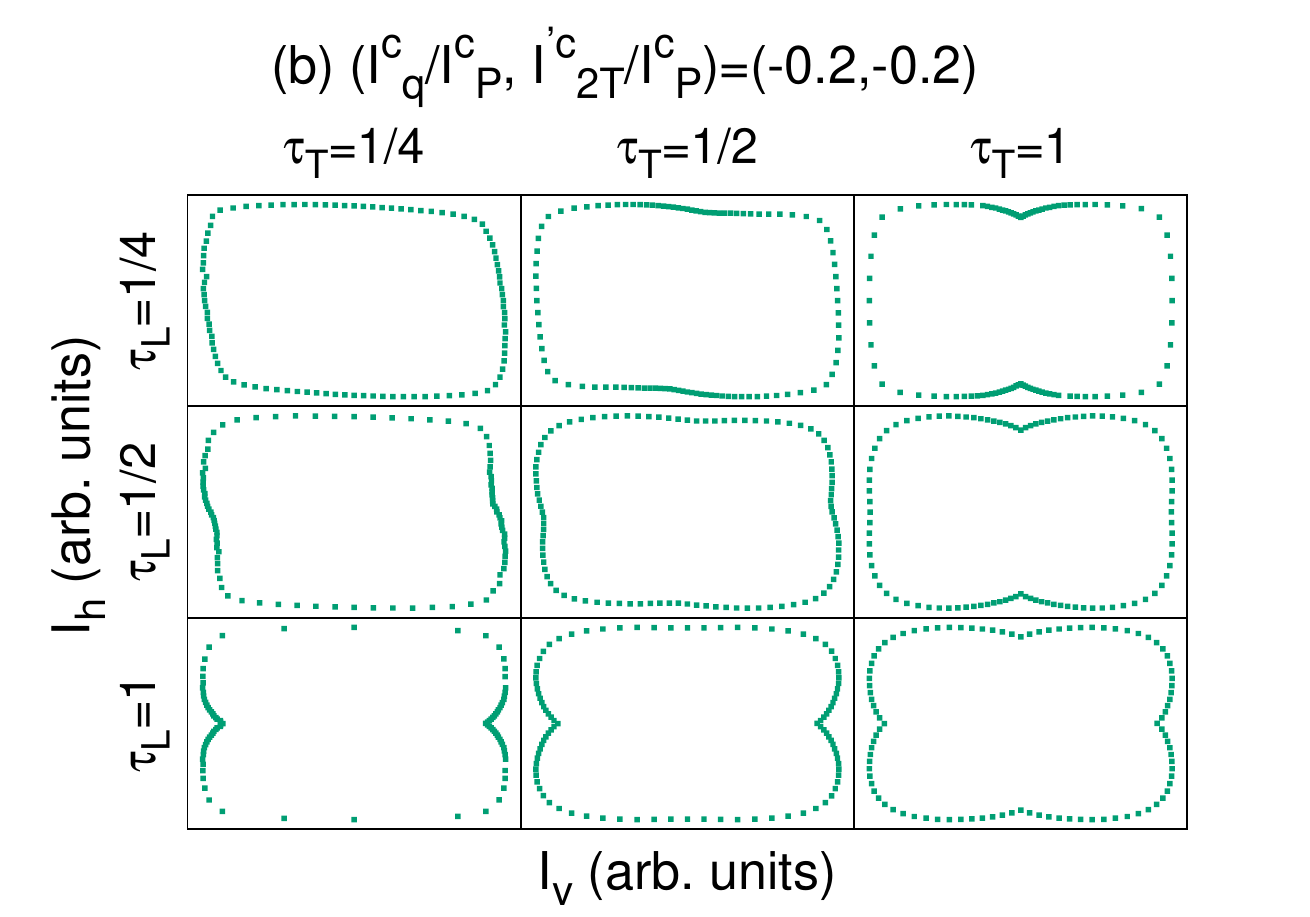}}
  \centerline{\includegraphics[width=.5\textwidth]{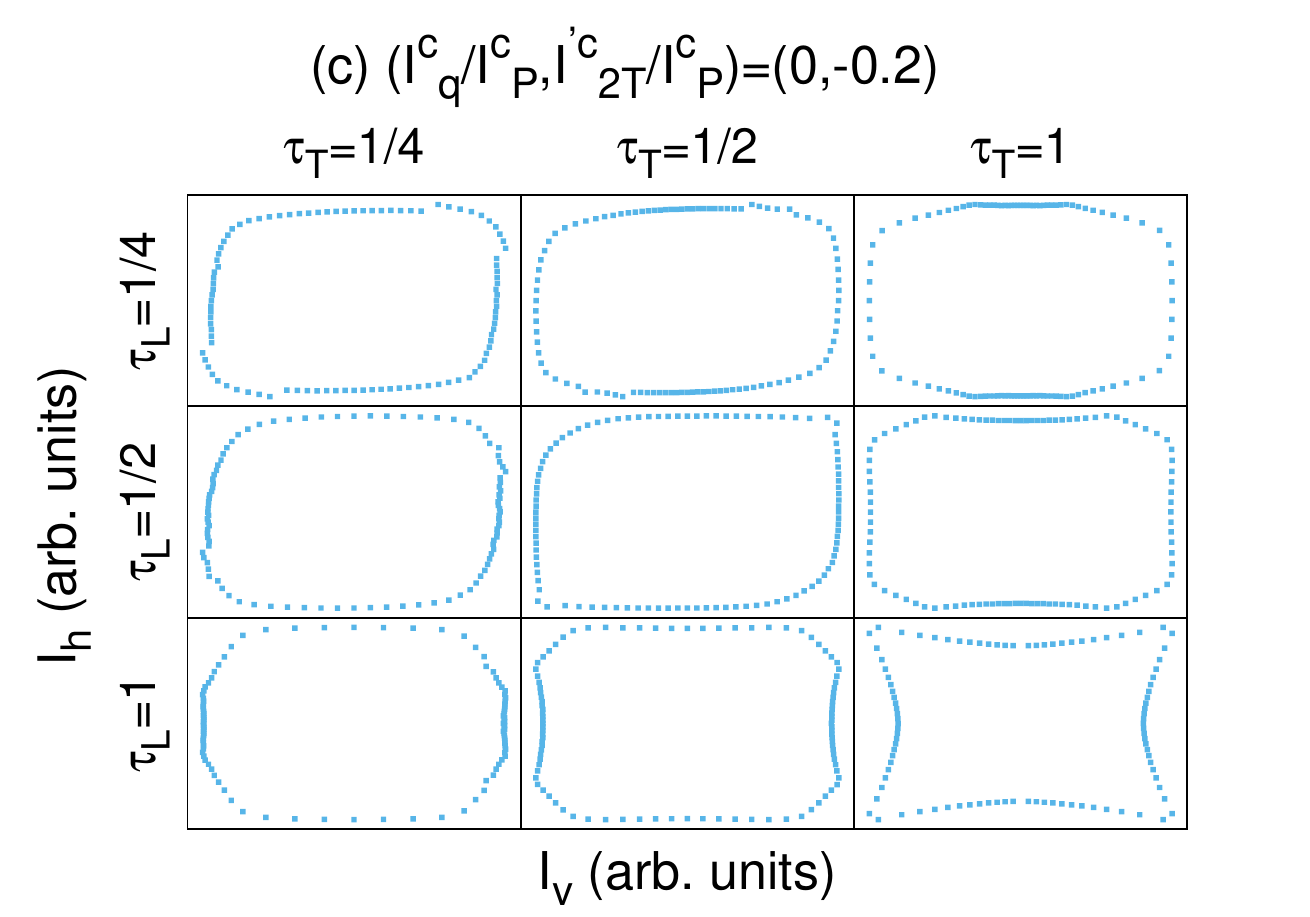}\includegraphics[width=.5\textwidth]{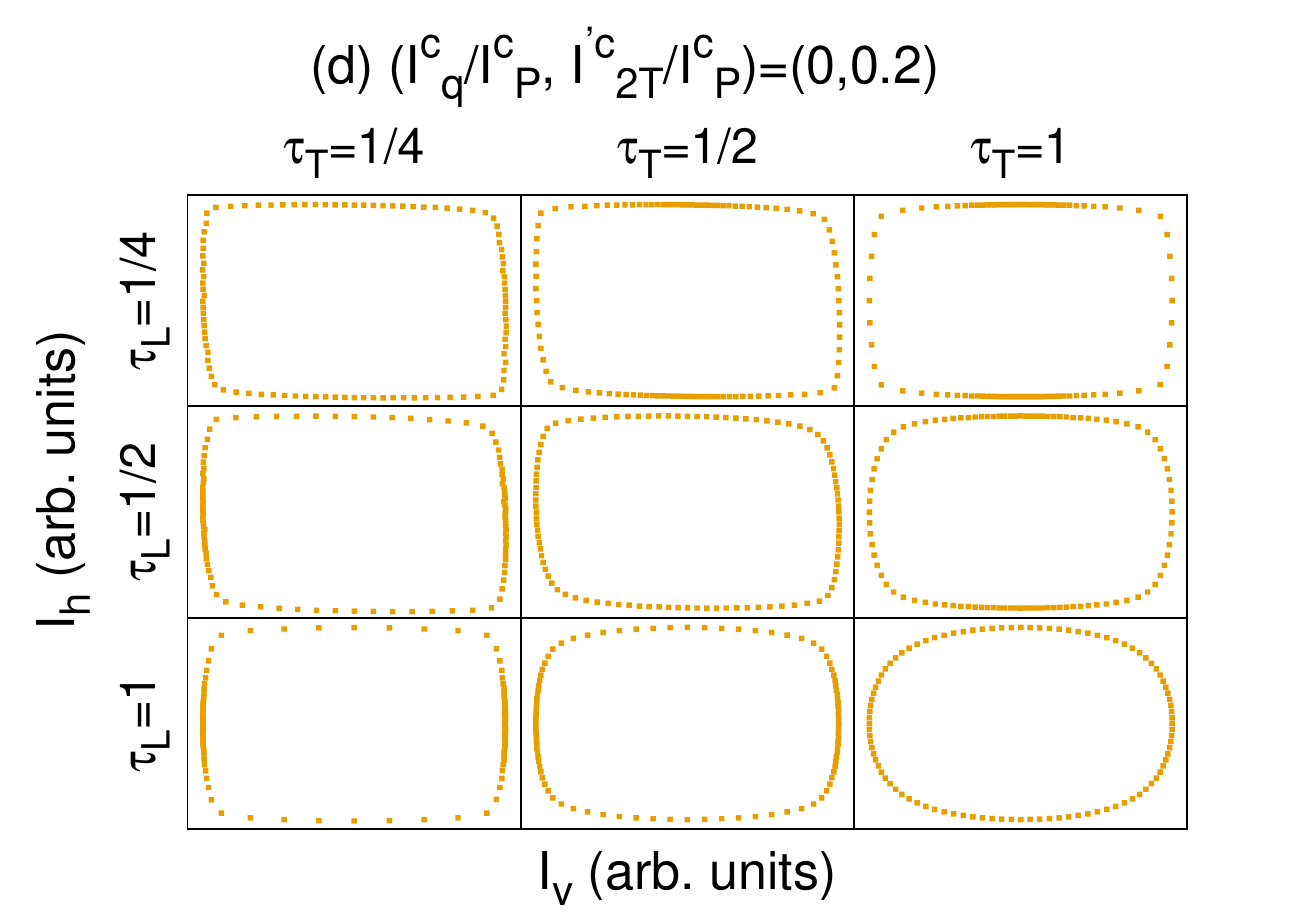}}
  \caption{Critical current contours in the
      $(I_v,I_h)$ plane, with $\tau_B=\tau_R=1$, and with two weak
      links. The values ($I_{q}^c/I_P^c,I^{'c}_{2T}/I_P^c)= (-0.2,0)$,
      $(-0.2,-0.2)$, $(0,-0.2)$ and $(0,0.2)$ are used on panels a-d
      respectively. The panels are organized as a table, and the
      values of $\tau_L$, $\tau_T$ are indicated. Each panel is
    rescaled to full size on the $I_{v}$ and
    $I_{h}$-axis. Polarization is with two orthogonal current biases,
    see Fig.~\ref{fig:thedevice}b. Temperature is
    set to zero.
  \label{fig:CCC_versus_ra_et_rc_xEJ_quartets_moins_0_2}
  }
\end{figure*}

\begin{figure}[htb]
  \centerline{\includegraphics[width=\columnwidth]{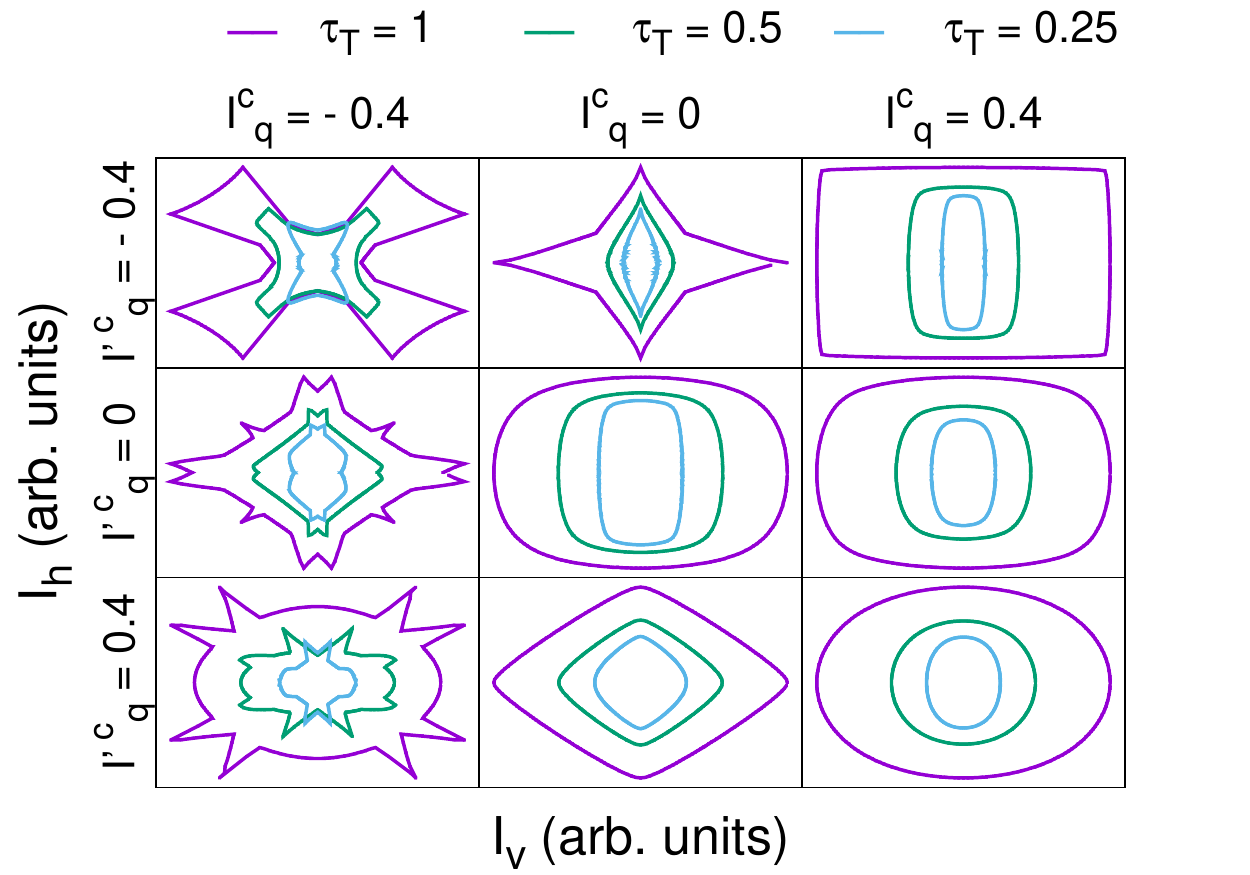}}
  \caption{Critical current contours in the $(I_v,I_h)$ plane, in the
    presence of Josephson coupling to the four superconducting phase
    variables, and with a single weak link. The contact transmission
    coefficients are such that $\tau_B=\tau_L=\tau_R=1$ and $\tau_T=1$
    (magenta), $\tau_T=0.5$ (green) and $\tau_T=0.25$ (blue). The
    different panels are organized like a table, and the values of
    $I_q^c$ and $I^{'c}_q$ are indicated on the figure, both of those
    being normalized to $I_P^c=1$. Each panel is rescaled to full size
    on the $I_{v}$ and $I_{h}$-axis. Polarization is with two
    orthogonal current biases, see
   Fig.~\ref{fig:thedevice}b. Temperature is set to
    zero.
    \label{fig:CCC2-final}
  }
\end{figure}

\subsection{Polarization with two orthogonal current biases}

In this subsection, we numerically solve
a related model where we impose current biases in both
horizontal and vertical directions, such that $I_{v}=I_T=-I_B$ and
$I_{h}=I_R=-I_L$ (see Fig.~\ref{fig:thedevice}b).
The four superconducting phase variables adjust accordingly. The
numerical calculations are based on evaluating convergence of the
steepest descent algorithm for a multiterminal Josephson junction. A
dichotomic search was implemented, in order to locate the CCCs to high
accuracy. We use $I^c_{k,l}\equiv I_P^c$ and $I^c_{k,l,(m)} \equiv
I_q^c$ for the critical currents of the processes coupling to two and
three superconducting phase variables,
respectively. Fig.~\ref{fig:CCC1} shows the CCCs of
a four-terminal device with the transmission coefficient scaling
factors $\tau_B=\tau_L=\tau_R=1$ and different values of $\tau_T$.
For positive values of $I_q^c/I_P^c$, the CCCs have the shape of
nested rounded rectangles. For sufficiently negative $I_q^c/I_P^c$
however, the CCCs evolve from diamond-like to a shape presenting
nonconvex sharp-angled points when lowering $\tau_T$.  Notably, the
CCCs with nonconvex sharp-angled points are only obtained for a
sufficiently negative Cooper quartet critical
  current (here $I_q^c/I_P^c=-0.2$), which is in
  agreement with the preceding analytical solution.

In
Fig.~\ref{fig:CCC_versus_ra_et_rc_xEJ_quartets_moins_0_2}a,
we further implement two weak links with $\tau_T,\,\tau_L \leq 1$,
while maintaining $\tau_B=\tau_R=1$, and we use a negative Cooper
quartet critical current 
$I_q^c/I_P^c=-0.2$. Focusing on the panels on the diagonal,
i.e. $\tau_T=\tau_L=1/4,\,1/2,\,1$, we obtain an evolution from
diamond-like to square-like CCCs, as $\tau_T=\tau_L$ decreases. Since
a rectangular CCC is indicative of independent currents in orthogonal
directions, this evolution demonstrates a loss of quantum mechanical
coupling between $I_{v}$ and $I_{h}$ as the contact transmission
coefficient scaling factor decreases. The intermediate value
$\tau_T=\tau_L=1/2$ yields reentrance on both supercurrent axes, which
originates from the underlying diagonal mirror symmetry in the
device. Considering now the off-diagonal panels in
Fig.~\ref{fig:CCC_versus_ra_et_rc_xEJ_quartets_moins_0_2}a,
we obtain shapes with nonconvex sharp-angled points on the $I_{v}^c$ axis if
$\tau_T=1/4,\,1/2$ and $\tau_L=1$, and the same on the $I_{h}^c$ axis if $\tau_T=1$ and
$\tau_L=1/4,\,1/2$. This is again in qualitative
  agreement with the analytical model calculations presented in the
  above Sec.~\ref{sec:mixt-pol}.

In Fig.~\ref{fig:CCC_versus_ra_et_rc_xEJ_quartets_moins_0_2}b, we
introduce all possible higher-order two-terminal
$I^{'c}_{2T}\sin(2(\varphi_i-\varphi_j))$ coupling terms, in addition
to the Cooper quartets. We observe the robustness of the reentrant
sharp-angled points with respect to addition of these. Qualitatively,
this can be interpreted as due to the fact that a smooth feature on
top of a sharp cusp does not alter the
latter. Figs.~\ref{fig:CCC_versus_ra_et_rc_xEJ_quartets_moins_0_2}c
and d comparatively show the CCCs with vanishingly small quartet
critical current but with finite $I^{'c}_{2T}$ taking negative or
positive values. The nonconvex sharp-angled points are absent in the
corresponding CCCs if $I^c_q=0$ and $I^{'c}_{2T}\ne 0$. Those
nonconvex sharp-angled points are thus a unique signature of the
nonlocally $\pi$-shifted Cooper quartets.

Eventually, we demonstrate robustness of the reentrant pockets upon
including the DC-Josephson effect depending on all four
superconducting phase variables \cite{split-quartets}. At the lowest
order in tunneling, the corresponding Josephson
critical currents are denoted by $I^{'c}_q$ and they
scale like $\tau_L \tau_R \tau_B \tau_T$.
Fig.~\ref{fig:CCC2-final} provides the CCCs for
variable combinations of $I_q^c/I_P^c$ and $I^{'c}_q/I_P^c$, and with
$\tau_T\alt 1$ and $\tau_B=\tau_L=\tau_R=1$. The data with $I_q^c\agt
0$ reveal smooth nonreentrant variations, contrasting with the sharper
reentrant-like variations on the other panels. We conclude that
reentrant features in CCCs at negative Cooper quartet
critical current $I_q^c$ are robust with respect to
including higher-order Josephson terms.

\section{Conclusions}
\label{sec:conclusions}

To conclude, it follows from basic theoretical arguments that the
quartet supercurrent contribution must be $\pi$-shifted with respect
to the lowest order Josephson Cooper pair supercurrent. We
demonstrated that the nonconvex two-dimensional critical current
contours (CCCs) of a current-biased four-terminal Josephson junction
are generically due to a relative $\pi$-shift of the higher-order
terms in the current-phase relation. These can either originate simply
from the two-terminal Josephson current-phase relation, or, more
interestingly, from the Cooper quartets. Finally, we demonstrated that
nonconvex sharp-angled points in the CCCs are a distinctive signature
of negative Cooper critical current contributions. However, we note
that too small negative quartet critical currents will restore convex
CCC, which sets constraints on the transmissions for the observation
of the characteristic reentrance. A recent experiment
\cite{multiterminal-exp3} reported the appearance of nonconvex CCCs
only under applied magnetic field. However, in contrast to our
assumptions, all contacts had large transparencies. Conclusive
evidence for the $\pi$-shifted quartet term could be realized with
bilayer graphene- or semiconducting-quantum point contacts
\cite{Danneau2,Pribiag} with tunable contact transparencies.

\section*{Acknowledgements}

The authors benefited from fruitful discussions with M. d'Astuto,
S. Collienne, T. Klein, F. L\'evy-Bertrand, M.A. M\'easson,
P. Rodi\`ere, and A. Silhanek.  R.M. acknowledges a useful
correspondence with V. E. Manucharyan.  R.M. thanks the Infrastructure
de Calcul Intensif et de Donn\'ees (GRICAD) for use of the resources
of the M\'esocentre de Calcul Intensif de l’Universit\'e
Grenoble-Alpes (CIMENT). This work was supported by the International
Research Project SUPRADEVMAT between CNRS in Grenoble and KIT in
Karlsruhe. This work received support from the French National
Research Agency (ANR) in the framework of the Graphmon
(ANR-19-CE47-0007) and JOSPEC (ANR-17-CE30-0030) projects. This work
was partly supported by Helmholtz Society through program STN and the
DFG via the Project No. DA 1280/7-1.

\appendix

\section{Details on phenomenologically squaring
  the single Cooper pair wave-function}
\label{sec:I}
In this Appendix, we detail how to deduce Eq.~(\ref{eq:double-split})
  from Eq.~(\ref{eq:single-split}). Namely, we square the
  wave-function of a Cooper pair split between $S_i$ and $S_j$:
\begin{eqnarray}
    \psi^2&=&\left[
    \frac{1}{\sqrt{2}}\left(c_{i,\uparrow}^+ c_{j,\downarrow}^+ -
    c_{i,\downarrow}^+ c_{j,\uparrow}^+\right)\right]^2\\
    &=& \frac{1}{2}
    \left(c_{i,\uparrow}^+ c_{j,\downarrow}^+ -
    c_{i,\downarrow}^+ c_{j,\uparrow}^+\right) \times \left(c_{i,\uparrow}^+ c_{j,\downarrow}^+ -
    c_{i,\downarrow}^+ c_{j,\uparrow}^+\right)\\
    &=& \frac{1}{2} \left[
    c_{i,\uparrow}^+ c_{j,\downarrow}^+c_{i,\uparrow}^+ c_{j,\downarrow}^+
    - c_{i,\uparrow}^+ c_{j,\downarrow}^+c_{i,\downarrow}^+ c_{j,\uparrow}^+\right.\\
    \nonumber
    &&\left.
    - c_{i,\downarrow}^+ c_{j,\uparrow}^+ c_{i,\uparrow}^+ c_{j,\downarrow}^+
    + c_{i,\downarrow}^+ c_{j,\uparrow}^+c_{i,\downarrow}^+ c_{j,\uparrow}^+ \right]\\
    \label{eq:prov1}
    &=& \frac{1}{2} \left[
      - \left( c_{i,\uparrow}^+ \right)^2\left(c_{j,\downarrow}^+\right)^2
      -\left(c_{i,\uparrow}^+c_{i,\downarrow}^+\right)\left(c_{j,\uparrow}^+c_{j,\downarrow}^+\right)
      \right.\\
      && \left.
      -\left(c_{i,\uparrow}^+c_{i,\downarrow}^+\right)\left(c_{j,\uparrow}^+c_{j,\downarrow}^+\right)
      - \left( c_{i,\downarrow}^+ \right)^2\left(c_{j,\uparrow}^+\right)^2\right]
    \nonumber
    .
\end{eqnarray}
Evaluating quantum mechanical expectation values in the final state
leads to
\begin{equation}
  \langle \psi^2 \rangle= - \langle
  c_{i,\uparrow}^+c_{i,\downarrow}^+\rangle \times \langle
  c_{j,\uparrow}^+c_{j,\downarrow}^+\rangle ,
    \label{eq:prov2}
\end{equation}
where we used $\langle \left(c_{i,\uparrow}^+\right)^2 \rangle=0$
because of spin conservation. The above Eq.~(\ref{eq:prov2}) matches
the above Eq.~(\ref{eq:double-split}).

\section{Details on a single superconducting weak link}
\label{sec:II}

In this Appendix, we evaluate the first- and second-order harmonics of
the Josephson current-phase relation for a single-channel
superconducting weak link having hopping amplitude $J_0$, and
connecting the ``left'' and ``right'' superconducting leads $S_L$ and
$S_R$ with superconducting phases $\varphi_L$ and $\varphi_R$
respectively. The Andreev Bound State (ABS) energies take the
following form:
\begin{equation}
  E_\pm(\varphi_R-\varphi_L,\alpha)=\pm \Delta
  \sqrt{1-\alpha \sin^2\left(\frac{\varphi_R-\varphi_L}{2}\right)}
  ,
\end{equation}
where the dimensionless normal-state transmission coefficient $\alpha$
between $0$ and $1$ is given by
\begin{equation}
  \alpha=\frac{4 (J_0/W)^2}{\left[1+(J_0/W)^2\right]^2}
 .
\end{equation}
The ABS energies are expressed as
\begin{eqnarray}
&&  E_\pm(\varphi_R-\varphi_L,\alpha)= \pm \Delta
  \sqrt{1-\frac{\alpha}{2} + \frac{\alpha}{2}
    \cos\left(\varphi_R-\varphi_L \right)}\\
  &&= \pm \Delta \sqrt{1-\frac{\alpha}{2}}
  \sqrt{1+\frac{\alpha}{2-\alpha} \cos\left(\varphi_R - \varphi_L\right)}
  .
\end{eqnarray}
Expanding to second order, we obtain
\begin{widetext}
\begin{eqnarray}
  E_\pm(\varphi_R-\varphi_L,\alpha)&=& \pm \Delta
  \sqrt{1-\frac{\alpha}{2}}
  \left\{1+\frac{\alpha}{2(2-\alpha)}\cos(\varphi_R-\varphi_L)
    -\frac{1}{8} \left(\frac{\alpha}{2-\alpha}\right)^2 \cos^2
    \left(\varphi_R-\varphi_L\right)\right\}
    + ...
\end{eqnarray}
\end{widetext}

Using $\cos^2\left(\varphi_R-\varphi_L\right)=
\left[1+\cos\left(2(\varphi_R-\varphi_L)\right)\right]/2$, we obtain
\begin{eqnarray}
E_\pm(\varphi_R-\varphi_L,\alpha)&=&\pm\left[E_0^J+E^J_{2e} \cos
  \left(\varphi_R-\varphi_L\right)\right.\\
  \nonumber
  &&+ \left.
  E^J_{4e} \cos\left(2\left(\varphi_R-\varphi_L\right)\right) + ...\right],
\end{eqnarray}
with
\begin{equation}
  \label{eq:limiting-behavior}
    \frac{E^J_{4e}}{E^J_{2e}}=-\frac{\alpha}{16}
\end{equation}
in the small-$\alpha$ limit, where $E^J_{2e}>0$ is positive and
$E^J_{4e}<0$ is negative.

\begin{figure}[htb]
  \includegraphics[width=\columnwidth]{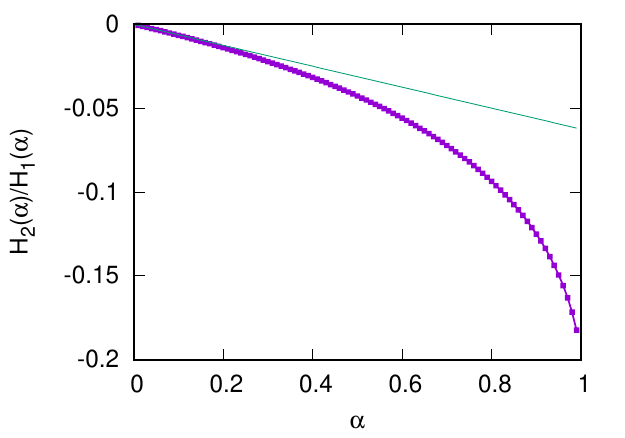}
  \caption{The figure shows $H_2(\alpha)/H_1(\alpha)$ as a function of
    the normal-state contact transparency $\alpha$ (magenta
    datapoints), where $H_1(\alpha)$ and $H_2(\alpha)$ are the first
    and second harmonics of the Josephson energy, see
    Eqs.~(\ref{eq:H1}) and~(\ref{eq:H2}). The green line shows a
    comparison to $H_2(\alpha)/H_1(\alpha)=-\alpha/16$, see
    Eq.~(\ref{eq:limiting-behavior}).
    \label{fig:H1-H2}
  }
\end{figure}

Now, we present supplemental numerical calculations for the amplitudes
$H_1(\alpha)$ and $H_2(\alpha)$ of the first and second Josephson
harmonics as a function of the dimensionless normal-state transmission
coefficient $\alpha$:
\begin{eqnarray}
  \label{eq:H1}
  H_1(\alpha)&=& \int_0^{2\pi} \frac{d \varphi}{2\pi} E_+(\varphi,\alpha) \cos\varphi\\
  \label{eq:H2}
H_2(\alpha)&=& \int_0^{2\pi} \frac{d \varphi}{2\pi} E_+(\varphi,\alpha) \cos(2\varphi)
.
\end{eqnarray}
Figure~\ref{fig:H1-H2} shows that $H_2(\alpha)/H_1(\alpha)<0$ is
negative for all values of $\alpha<1$. The limiting behavior
$H_2(\alpha) /H_1(\alpha)=-\alpha/16$ is obtained at small~$\alpha$,
in agreement with the above Eq.~(\ref{eq:limiting-behavior}).

\section{Details on the device controlled with one current and one
phase bias}
\label{sec:III}
In this Appendix, we consider the multiterminal Josephson circuit
shown on the above figure~\ref{fig:thedevice}a, consisting of the four
superconducting leads $S_L$, $S_R$, $S_B$ and $S_T$ with the phases
$\varphi_L$, $\varphi_R$, $\varphi_B$ and $\varphi_T$. The phase
difference $\varphi_R-\varphi_L=\Phi$ is controlled by the flux $\Phi$
in the loop, and supercurrent $I_{v}=I_B=-I_T$ is forced to flow from
the ``bottom'' to the ``top'' superconducting leads.  We use the
notation $\varphi_T=\psi$. Given those constraints, the supercurrent
transmitted into $S_T$ is parameterized by a single phase variable,
for instance by the phase variable $\psi$, and the critical current
from bottom to top is obtained by maximizing the supercurrent
$I_{v}(\psi)$ over $\psi$.

The superconductor $S_T$ is assumed to be connected to the normal
conductor $N_0$ by hopping amplitude that is weaker than the others.
A reduction factor $\tau_T$ is applied to each Cooper pair crossing
the $N_0-S_T$ interface.

\begin{figure}[htb]
  \includegraphics[width=\columnwidth]{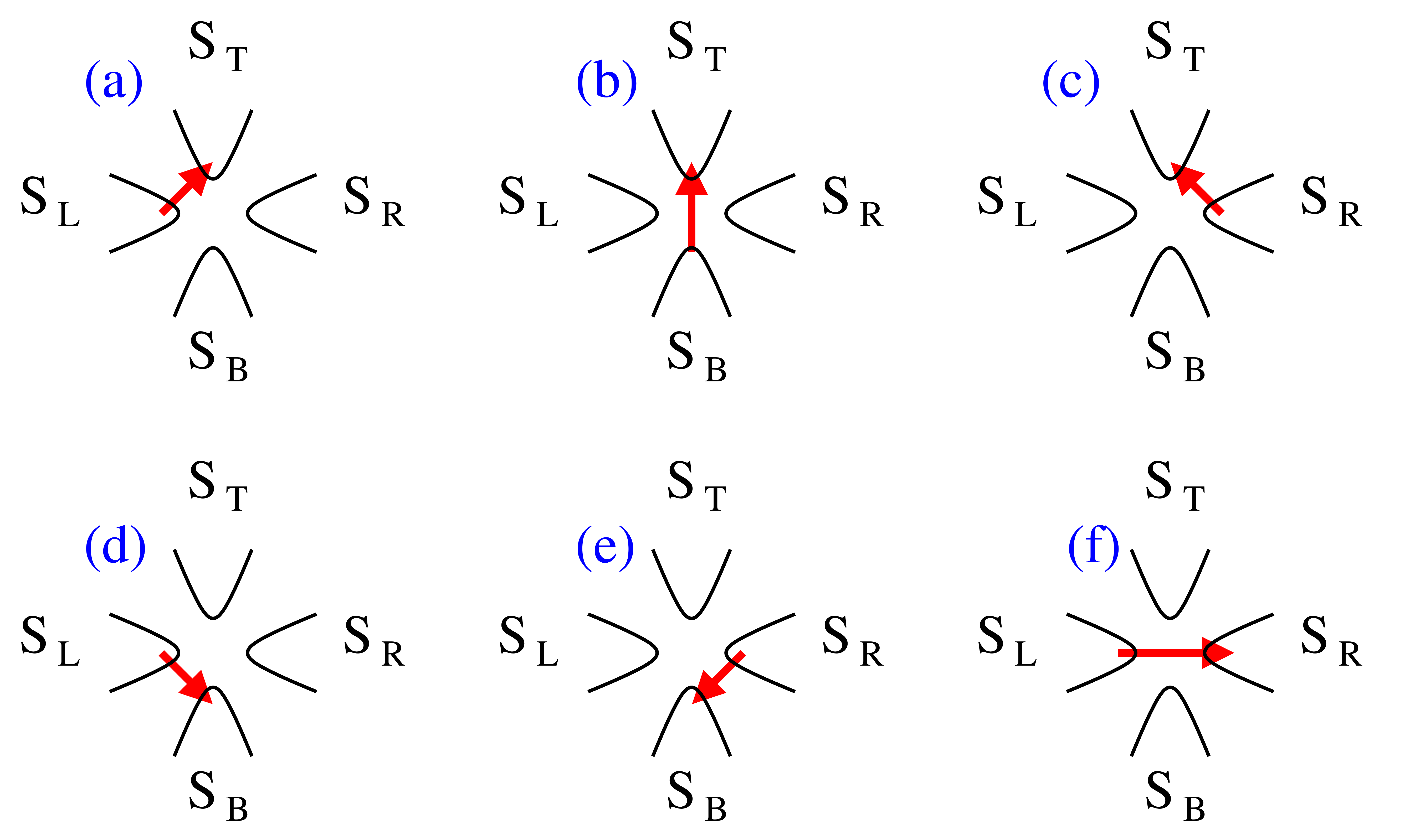}
  \caption{The figure shows the six two-terminal processes transferring
    Cooper pairs between the superconducting leads.
    \label{fig:2T}
  }
\end{figure}

First considering vanishing small quartet Josephson energy $I^c_q=0$, we
obtain the following expression of the four-terminal Josephson
junction energy $E^{(0)}$:
\begin{eqnarray}
\label{eq:sum1-0}
  E^{(0)}&=&
  E_P \tau_T \left\{\cos\left(\varphi_T-\varphi_L\right)
  + \cos\left(\varphi_T-\varphi_B\right)\right.\\
  \nonumber
  &&  +\left. \cos\left(\varphi_T-\varphi_R\right)\right\}\\
  \nonumber
  &+&
  E_P \left\{\cos\left(\varphi_B-\varphi_L\right)
  + \cos\left(\varphi_B-\varphi_R\right)\right.\\
  \nonumber
  &&
  + \left. \cos\left(\varphi_R-\varphi_L\right)\right\}
  ,
\end{eqnarray}
where $E_P$ is the Josephson energy associated to transferring Cooper
pairs between the leads. Each term entering Eq.~(\ref{eq:sum1-0}) is
schematically shown in figure~\ref{fig:2T}. Using
$\varphi_L=-\varphi_R=-\varphi/2$ and $\varphi_T=\psi$, we obtain
\begin{eqnarray}
\nonumber
  E^{(0)}&=&
  E_P \tau_T \left\{\cos\left(\psi+\frac{\varphi}{2}\right)
  + \cos\left(\psi-\varphi_B\right)
  + \cos\left(\psi-\frac{\varphi}{2}\right)\right\}\\
&&  +
  E_P \left\{\cos\left(\varphi_B+\frac{\varphi}{2}\right)
  + \cos\left(\varphi_B-\frac{\varphi}{2}\right)
  + \cos\varphi\right\}
  .\label{eq:sum1}
\end{eqnarray}
Then:
\begin{eqnarray}
  I_T&=&-\frac{2e}{\hbar}\frac{\partial E^{(0)}}{\partial \psi}\\
  &=&
  \frac{2eE_P \tau_T}{\hbar} \left\{
  \sin\left(\psi+\frac{\varphi}{2}\right)
  +
  \sin\left(\psi-\varphi_B\right)\right.\\
  \nonumber
&&\left.  +
  \sin\left(\psi-\frac{\varphi}{2}\right)\right\}\\
I_B&=&-\frac{2e}{\hbar}\frac{\partial E^{(0)}}{\partial \varphi_B}\\
  &=&
  \frac{2eE_P \tau_T}{\hbar}
  \sin\left(\varphi_B-\psi\right)\\
  &&+ \frac{2E_P}{\hbar}\left\{
  \sin\left(\varphi_B+\frac{\varphi}{2}\right)
  +
  \sin\left(\varphi_B-\frac{\varphi}{2}\right)\right\}
  .\nonumber
\end{eqnarray}
\begin{figure}[htb]
  \includegraphics[width=\columnwidth]{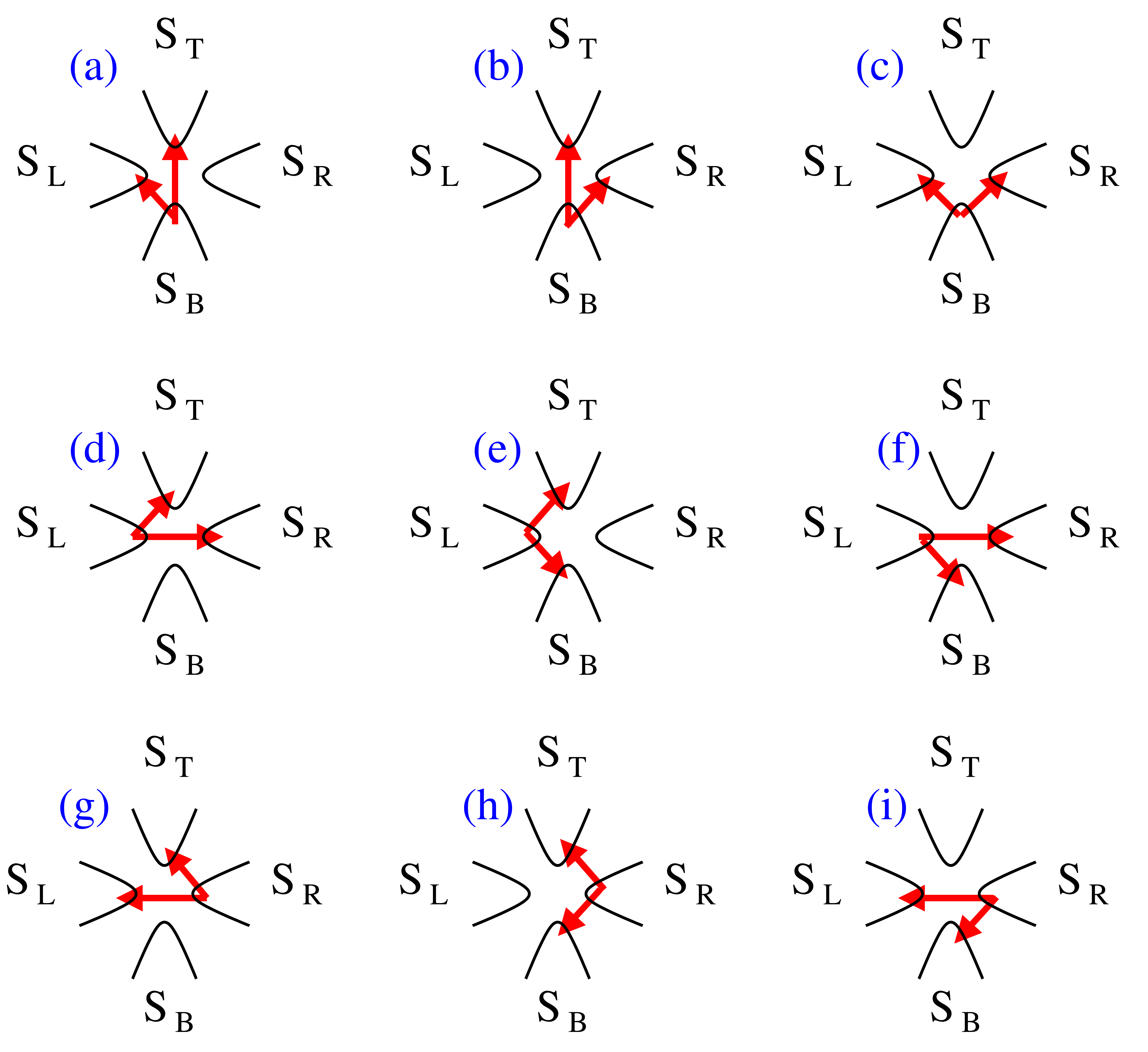}
  \caption{The figure shows the nine three-terminal processes
    transferring Cooper quartets between the superconducting leads, at
    the orders $(\tau_T)^0$ and $\tau_T$. The three higher-order
    processes of order $(\tau_T)^2$ are not shown on the figure.
    \label{fig:3T}
  }
\end{figure}
The current source imposes
\begin{equation}
I_T+I_B=-(2e/\hbar) \left[ \partial
  E^{(0)}/\partial \psi + \partial E^{(0)}/\partial
  \varphi_B\right]=0,
\end{equation}
which leads to the self-consistent
$\varphi_B=\varphi_B^*$, with
\begin{equation}
  \label{eq:phi-B-*}
  \sin\varphi_B^*=-\tau_T \sin\psi
  ,
\end{equation}
showing that $\varphi_B^*$ is of order $\tau_T$. Then, at the order
$\tau_T$, we obtain
\begin{equation}
  \label{eq:toto1}
  I_T=-\frac{2e}{\hbar}\frac{\partial E^{(0)}}{\partial \psi}
  \simeq
  \frac{2eE_P \tau_T}{\hbar} \left[1+2\cos\left(\frac{\varphi}{2}\right)\right]\sin\psi
  .
\end{equation}
Taking the maximum over $\psi$ and expanding in small $\varphi$ leads
to following expansion of the critical current flowing from bottom to
top at small $\varphi$, at the order $\tau_T$:
\begin{equation}
  \label{eq:Ic-convex}
  I_c\simeq \frac{2eE_P \tau_T}{\hbar} \left[3-\frac{\varphi^2}{4}\right]
  ,
\end{equation}
leading to convex CCC in absence of the quartets.
Eq.~(\ref{eq:Ic-convex}) is rewritten as
\begin{equation}
  \label{eq:Ic-convex-2}
  I_c\simeq \tau_T I_P^c \left[3-\frac{\varphi^2}{4}\right]
  ,
\end{equation}
where $I_P^c=2e E_P/\hbar$ is related to the Cooper pair critical current.

Now, we include finite but small quartet Josephson energy $E_q$. Those
processes are shown in figure~\ref{fig:3T}, and, at the order
$\tau_T$, they yield the following correction $\delta E^{(0)}$ to the
energy $E^{(0)}$ in Eq.~(\ref{eq:sum1-0}):
\begin{widetext}
\begin{eqnarray}
  \delta E^{(0)}&=& E_q \left\{
  \tau_T \cos\left(\varphi_L+\varphi_T-2\varphi_B\right)
+ \tau_T \cos\left(\varphi_R+\varphi_T-2\varphi_B\right)
+ \cos\left(\varphi_R+\varphi_L-2\varphi_B\right)
+\tau_T \cos\left(\varphi_T+\varphi_R-2\varphi_L\right)\right.\\
\nonumber
&&+\left.\tau_T \cos\left(\varphi_T+\varphi_B-2\varphi_L\right)
+ \cos\left(\varphi_R+\varphi_B-2\varphi_L\right)
+\tau_T \cos\left(\varphi_T+\varphi_L-2\varphi_R\right)
+\tau_T \cos\left(\varphi_T+\varphi_B-2\varphi_R\right)\right.\\
\nonumber
&&+ \left.\cos\left(\varphi_L+\varphi_B-2\varphi_R\right)
\right\}
.
\end{eqnarray}
Then, we find
\begin{eqnarray}
  \label{eq:1}
-  \frac{\partial\delta E^{(0)}}{\partial \varphi_T}&=& \tau_T E_q \left\{
  \sin\left(\varphi_L+\varphi_T-2\varphi_B\right)
+ \sin\left(\varphi_R+\varphi_T-2\varphi_B\right)
+\sin\left(\varphi_T+\varphi_R-2\varphi_L\right)
+ \sin\left(\varphi_T+\varphi_B-2\varphi_L\right)\right.\\
\nonumber
&&+ \left.\sin\left(\varphi_T+\varphi_L-2\varphi_R\right)
+ \sin\left(\varphi_T+\varphi_B-2\varphi_R\right)\right\}\\
\label{eq:1-bis}
&=&\tau_T E_q \left\{
\sin\left(-\frac{\varphi}{2}+\psi-2\varphi_B\right)
+ \sin\left(\frac{\varphi}{2}+\psi-2\varphi_B\right)
+\sin\left(\psi+\frac{3\varphi}{2}\right)
+ \sin\left(\psi+\varphi_B+\varphi\right)\right.\\
\nonumber
&&+\left. \sin\left(\psi-\frac{3\varphi}{2}\right)
+ \sin\left(\psi+\varphi_B-\varphi\right)\right\}
.
\end{eqnarray}
Since $\varphi_B$ is of order $\tau_T$ [see the above
  Eq.~(\ref{eq:phi-B-*})], we replace $\varphi_B$ by $\varphi_B=0$ in
the above Eqs.~(\ref{eq:1})-(\ref{eq:1-bis}) to obtain the
order-$\tau_T$ contribution to $-\partial\delta E^{(0)}/\partial
\varphi_T$:
\begin{eqnarray}
- \frac{\partial\delta E^{(0)}}{\partial \varphi_T}&\simeq& \tau_T
E_q \left\{ \sin\left(-\frac{\varphi}{2}+\psi\right) +
\sin\left(\frac{\varphi}{2}+\psi\right)
+\sin\left(\psi+\frac{3\varphi}{2}\right)+
\sin\left(\psi+\varphi\right) +
\sin\left(\psi-\frac{3\varphi}{2}\right) +
\sin\left(\psi-\varphi\right)\right\}\\ &=& 2\tau_T E_q \left\{
\cos\left(\frac{\varphi}{2}\right) +\cos\varphi
+\cos\left(\frac{3\varphi}{2}\right)\right\} \sin\psi .
\label{eq:toto2}
\end{eqnarray}
Considering now $-\partial\delta E^{(0)}/\partial \varphi_B$, we find
\begin{eqnarray}
- \frac{\partial\delta E^{(0)}}{\partial \varphi_B}&=& E_q \left\{ 2
\tau_T \sin\left(2\varphi_B-\varphi_L-\varphi_T\right) + 2 \tau_T
\sin\left(2\varphi_B-\varphi_R-\varphi_T\right)+2
\sin\left(2\varphi_B-\varphi_R-\varphi_L\right) +\tau_T
\sin\left(\varphi_T+\varphi_B-2\varphi_L\right)\right.\\\nonumber
&&+\left. \sin\left(\varphi_R+\varphi_B-2\varphi_L\right) +\tau_T
\sin\left(\varphi_T+\varphi_B-2\varphi_R\right)
+\sin\left(\varphi_L+\varphi_B-2\varphi_R\right) \right\}\\ &=&E_q
\left\{ 2 \tau_T \sin\left(2\varphi_B+\frac{\varphi}{2}-\psi\right) +
2 \tau_T \sin\left(2\varphi_B-\frac{\varphi}{2}-\psi\right) + 2
\sin\left(2\varphi_B\right)+\tau_T
\sin\left(\psi+\varphi_B+\varphi\right) \right.\\\nonumber &&+\left.
\sin\left(\varphi_B+\frac{3\varphi}{2}\right) +\tau_T
\sin\left(\psi+\varphi_B-\varphi\right) +\sin\left(\varphi_B-\frac{3
  \varphi}{2}\right) \right\} .
\end{eqnarray}
\end{widetext}

The phase variable $\varphi_B$ turns out to be linear in $\tau_T$ at
$E_q=0$ [see the above Eq.~(\ref{eq:phi-B-*})]. At small $E_q$, both
$-{\partial\delta E^{(0)}}/{\partial \varphi_T}$ and $-{\partial\delta
  E^{(0)}}/{\partial \varphi_B}$ are of order $E_q \tau_T$. Then,
$\varphi_B^*$ is linear in $\tau_T$ in the presence of small $E_q$, as
it was the case for $E_q=0$.

The supercurrent transmitted into the superconducting lead $S_T$ is then
given by the sum of Eqs.~(\ref{eq:toto1}) and~(\ref{eq:toto2}):
\begin{eqnarray}
  I_T&=&-\frac{2e}{\hbar}
  \frac{\partial \left(E^{(0)}+\delta E^{(0)}\right)}{\partial \psi}\\
 &\simeq& \frac{2e\tau_T E_P}{\hbar}
  \left\{3+6\frac{E_q}{E_P}
  - \frac{\varphi^2}{4} \left[ 1+14\frac{E_q}{E_P}\right]\right\} \sin\psi
  .
\end{eqnarray}
Taking the maximum over $\psi$ leads to the following expression of
the critical current:
\begin{eqnarray}
  \label{eq:provA}
  I_{v}^c&=&-\frac{2e}{\hbar}
  \frac{\partial \left(E^{(0)}+\delta E^{(0)}\right)}{\partial \psi}\\
  &\simeq& \frac{2e\tau_T E_P}{\hbar}
  \left\{3+6\frac{E_q}{E_P}
  - \frac{\varphi^2}{4} \left[ 1+14\frac{E_q}{E_P}\right]\right\}
  .
\end{eqnarray}
Eq.~(\ref{eq:provA}) is rewritten as
\begin{equation}
  \label{eq:provA-2}
  I_{v}^c=\simeq \tau_T I_P^c
  \left\{3+6\frac{I_q^c}{I_P^c}
  - \frac{\varphi^2}{4} \left[ 1+14\frac{I_q^c}{I_P^c}\right]\right\}
  ,
\end{equation}
where $I_P^c=2e E_P/\hbar$ and $I_q^c=2e E_q/\hbar$ are related to the
Cooper pair and Cooper quartet critical currents. This concludes the
demonstration of the above
Eq.~(\ref{eq:result}). Eq.~(\ref{eq:provA-2}) goes to
Eq.~(\ref{eq:Ic-convex-2}) in the $I_q^c\rightarrow 0$ limit of
vanishingly small quartet energy. As a consequence of
Eq.~(\ref{eq:provA-2}), the CCC is nonconvex if
\begin{equation}
  I^c_q<-\frac{I^c_P}{14}
  ,
\end{equation}
thus necessarily being negative.

\end{document}